\documentclass[11pt,a4paper]{article}
\pdfoutput=1

\usepackage{comment}
\usepackage{jheppub}
\usepackage{latexsym}
\usepackage{multirow}
\usepackage{enumerate}
\usepackage{color}
\usepackage[usenames,dvipsnames,table]{xcolor}
\usepackage{graphicx}
\usepackage{epsfig}  
\usepackage{epsf}    
\usepackage{dcolumn}
\usepackage{bm}
\usepackage{dcolumn}
\usepackage{textcomp}
\usepackage{float}
\usepackage{subfig}
\usepackage{hypcap}
\usepackage[]{hyperref}
\usepackage{makecell}
\usepackage{pifont}
\usepackage{appendix}
\usepackage{url}
\usepackage{amsmath}
\hypersetup{
  bookmarks=true,         
  unicode=false,          
  pdftoolbar=true,        
 pdfmenubar=true,        
 pdffitwindow=true,     
 pdfstartview={FitH},    
 pdfsubject={Neutrino Oscillations Phenomenology},   
 pdfnewwindow=true,      
 pdfcreator={RevTeX},
 colorlinks=true,       
 linkcolor=red,          
 citecolor=blue,        
 filecolor=black,      
 urlcolor=blue,           
  }

\preprint{TIFR/TH/21-12}

\title{Frugal $U(1)_X$ models with non-minimal flavor violation
  for $b \to s \ell \ell$ anomalies and neutrino mixing}

\author{Disha Bhatia$^{\;a}$,} 
\author{Nishita Desai$^{\;b}$,}
\author{Amol Dighe$^{\;b}$} 

\affiliation{$^{a}$Institute of Mathematical Sciences,
  CIT Campus, Taramani, Chennai 600113, India\\
  $^{b}$Tata Institute of Fundamental Research, Homi Bhabha Road, Colaba,
  Mumbai 400005, India}

\emailAdd{dishabhatia@imsc.res.in}
\emailAdd{nishita.desai@tifr.res.in }
\emailAdd{amol@theory.tifr.res.in}

\abstract{
  We analyze the class of models with an extra $U(1)_X$ gauge symmetry that 
  can account for the  $b \to s \ell \ell$  anomalies by modifying the Wilson
  coefficients $C_{9e}$ and $C_{9\mu}$ from their standard model values. 
  At the same time, these models generate appropriate quark mixing, and
  give rise to neutrino mixing via the Type-I seesaw mechanism.
  Apart from the gauge boson $Z'$, these frugal models only have three
  right-handed neutrinos for the seesaw mechanism, an additional $SU(2)_L$
  scalar doublet for quark mixing, and a SM-singlet scalar
  that breaks the $U(1)_X$ symmetry.
  This set-up identifies a class of leptonic symmetries, and necessitates
  non-zero but equal charges for the first two quark generations.
  If the quark mixing beyond the standard model were CKM-like, all
    these symmetries would be ruled out by the latest flavor constraints
    on Wilson coefficients and collider constraints on $Z'$ parameters.
  However, we identify a single-parameter source of non-minimal flavor 
  violation that allows a wider class of $U(1)_X$ symmetries to be compatible
  with all data.
  We show that the viable leptonic symmetries have to be of the form
   $L_e \pm 3 L_\mu - L_\tau$ or $L_e - 3 L_\mu + L_\tau$,
  and determine the $(M_{Z^\prime}, g_{Z^\prime})$
  parameter space that may be probed by the high-luminosity data at the LHC.
  }


\begin{document}
\maketitle
\flushbottom

\section{Introduction}
\label{sec:intro}

The flavor anomalies in the neutral-current transitions of several $b \to s$
processes have persisted for a long time~\cite{flavor-anomalies-review}. 
Among them, the observables falling under the class of generic ``ratio''
observables, i.e. $R_H \equiv B(B \to H \mu \mu)/B(B \to H e e)$ where
$H = K, K^*, X_S, \dots$, serve as gold standards for pointing to the existence
of lepton flavor universality violating (LFUV) new physics (NP)~\cite{RH-obs},
owing to their small theoretical uncertainties.
The values of these observables are close to unity in the standard model (SM),
for the carefully chosen di-lepton invariant mass-squared ($q^2$) bins.
These values are known to a great accuracy since the dominant theoretical
uncertainties from QCD largely cancel out in the ratio,
while the QED uncertainties
lead to only $\mathcal{O}(1\%)$ error in $R_{K^{(*)}}$ predictions~\cite{rk-qed-sm}.
In the SM, lepton flavor universality (LFU) is violated only
by the Higgs interactions, but since the relevant couplings are proportional
to lepton masses, the effect is too minuscule to make any difference to $R_H$.

The recent update on
$R_K= 0.846^{+0.042}_{-0.039}\text{(stat)} ^{+0.013}_{-0.012} \text{(syst)}$, 
measured in the $q^2$-bin $[1.1,6]$ GeV$^2$ 
by the LHCb collaboration~\cite{rk-2021}, is 3.1$\sigma$ away 
from the SM expectation $R_K^{\rm SM} = 1.0 \pm 0.01$\cite{rk-qed-sm},
and has strengthened the case for LFU violation.
This latest measurement is consistent with the previous measurements of
$R_K$~\cite{rk-2014,rk-2019}.
The LHCb measurements of another closely related ratio observable, $R_{K^*}$,
show a deviation from the SM predictions in the
low-$q^2$ ([0.04, 1.1] GeV$^2$) and
central-$q^2$ ([1.1, 6.0] GeV$^2$) bins~\cite{rks-2018}. 
There is expected to be a strong correlation between the NP contribution
to $R_K$ and the central-$q^2$ bin value of $R_{K^*}$.

There are also other $b \to s \mu \mu$ measurements which deviate from their SM
expectations at the $2\sigma-2.5\sigma$ level accuracy, for example,
the angular observable $P_5^\prime$ in $B^{0} \to K^{*0} \mu^+ \mu^-$~\cite{
  angular-Ks0-2021,angular-ATLAS,angular-Belle} and
$B^{+} \to K^{*+} \mu^+ \mu^-$~\cite{angular-Ks+-2020} channels,
and the branching ratio of $B_s \to \phi \mu^+ \mu^-$~\cite{Bstophimumu}
which is smaller than the SM expectation. 
Note that these measurements are not entirely free from hadronic uncertainties,
like the form factor uncertainties in the branching ratio observables, and the
non-factorizable contributions~\cite{charm-loop-Mannel,power-corr-matias}
due to charm loops in both branching ratio observables and $P_5^\prime$.
However, all these neutral current anomalies in combination point towards
LFUV new physics with more than $4\sigma$
significance. The exact quantification of the deviation of SM depends on the
method of combining data from different observations,
and assumptions on the power corrections~\cite{isidori-2021,shireen-2021,
  grienstein-2021,Altmannshofer-2021,
  matias-2021,mahmoudi-2021,matias-2019, mahmoudi-2019}.
In the coming years, the combined measurements from both Belle2 and LHC
are expected to shed more light on these anomalies~\cite{future-prospects}.

The effective field theory approach allows incorporating NP in $b \to s \ell \ell$
transitions in a model-independent manner, in the language of effective
higher-dimensional operators and their Wilson coefficients (WCs)~\cite{buras-review}. 
Global fits to the radiative, semileptonic, and leptonic $b \to s$ data~\cite{
  Altmannshofer-2021,grienstein-2021,shireen-2021,mahmoudi-2021, matias-2021}
indicate the extent of NP contributions to relevant combinations of WCs,
needed to account for the above neutral-current flavor anomalies.
It is observed that most of these anomalies may be explained by the NP
contributions to the vector and axial-vector $b \to s \ell \ell$ effective
operators
\begin{equation}
  \mathcal{O}_{9\ell}^{(\prime)} = \frac{\alpha_e}{4 \pi}\left[ \bar{s}\gamma_{\mu} P_{L(R)} b\right] \left[ \bar{\ell}\gamma^{\mu}  \ell\right] \quad \mbox{and} \quad 
  \mathcal{O}_{10 \ell}^{(\prime)} =\frac{\alpha_e}{4 \pi}\left[ \bar{s}\gamma_{\mu} P_{L(R)} b\right]
 \left[ \bar{\ell}\gamma^{\mu}  \gamma_5 \ell\right] \;,
 \label{eqn:axial-vector}
\end{equation}
whose WCs are denoted by $C_9^{(\prime)}$ and $C_{10}^{(\prime)}$, respectively. 
NP contributions to scalar/ pseudoscalar and tensor operators, though possible
in principle, do not lead to simultaneous explanations of multiple anomalies
in one-dimensional fits~\cite{diptimoy-tensor,Hiller-eft}.
The former also get stringent constraints from the $B_s\to \mu^+ \mu^-$ 
measurements which are in good agreement with the SM~\cite{Hiller-eft}.
 
Most of the anomalies discussed above involve muons, with the LFUV ratios
$R_{K^{(*)}}$ involving electrons in addition.
In order to keep the NP parameters to a minimum,
most of the global fits have been performed with the assumption of NP
only in the muon sector~\cite{Altmannshofer-2021,grienstein-2021,shireen-2021},
i.e., in terms of operators $\mathcal{O}_{9\mu}^{(\prime)}$ and
$\mathcal{O}_{10\mu}^{(\prime)}$ in the language of eq.~(\ref{eqn:axial-vector}). 
Since $R_{K^{(*)}}$ is observed to be less than its SM expectation, the NP 
effects are expected to be destructively interfering with the SM.
While one-dimensional fits~\cite{isidori-2021,mahmoudi-2021, matias-2021, shireen-2021,Altmannshofer-2021,grienstein-2021} 
prefer NP contributions to the WC combinations $C_{9\mu}^{\text{NP}}$,
$C_{9\mu}^\text{NP} = -C_{10\mu}^\text{NP}$, or
$C_{9\mu}^\text{NP} = -C_{9/10\mu}^{\prime}$,
the two-dimensional fits~\cite{isidori-2021,mahmoudi-2021, matias-2021, shireen-2021,Altmannshofer-2021,grienstein-2021}
favour new physics effects in the planes of the WC-pairs
$\left( C_{9\mu}^\text{NP}, C_{10\mu}^\text{NP}\right)$,
$\left( C_{9\mu}^\text{NP}, C_{9\mu}^{\prime}\right)$ 
and $\left( C_{9\mu}^\text{NP}, C_{10\mu}^{\prime}\right)$.
Note that for the WCs where the SM contribution is nonzero,
viz. $C_{9\ell}$ and $C_{10\ell}$, we denote the NP contribution as
$C_{9\ell}^{\rm NP}$ and $C_{10\ell}^{\rm NP}$, respectively.
For the primed operators, there is no SM contribution, and hence no
need to distinguish the NP contribution from the total one.

\begin{figure} [t]
\begin{center}
\includegraphics[width=0.45 \textwidth]{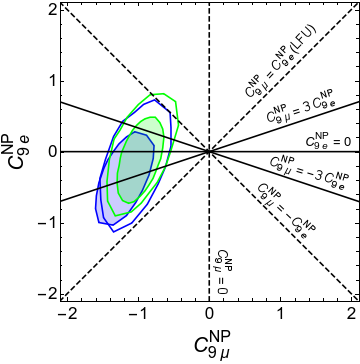}
\caption{The values in the ($C_{9 \mu}^{\text{NP}}$,$C_{9e}^{\text{NP}}$) plane,
  preferred at $2\sigma$ from global fits.
  The blue contours correspond to the fits in~\cite{matias-2021, matias-2019},
  and the green contours to the fits in~\cite{mahmoudi-2021, mahmoudi-2019}.
  The filled contours correspond to the fits based on the updates in
  Moriond 2021~\cite{matias-2021,mahmoudi-2021}, while the unfilled ones correspond
  to the older results
  based on data available after Moriond 2019~\cite{matias-2019,mahmoudi-2019}. 
  The black solid and dashed lines correspond to fixed ratios of
  $C_{9e}^\text{NP}$ and $C_{9\mu}^\text{NP}$. The ratios corresponding 
  to the dashed lines are disfavoured from the $b \to s$ global fits. }
       \label{fig:1}
       \end{center}
       \end{figure}

Although the involvement of NP in the muon sector is necessary to explain
the anomalies, it is quite possible that NP affects the electron sector also.
Recent global fits that take this into account~\cite{mahmoudi-2021, matias-2021}
indicate that the scenario with NP affecting
$\left( C_{9e}^\text{NP}, C_{9\mu}^\text{NP}\right)$ can also
explain the neutral-current flavor anomalies and other $b \to s$ measurements
reasonably well.
These fits are shown in figure~\ref{fig:1}.
The best-fit solution necessitates a negative value for $C_{9\mu}^\text{NP}$
in order to achieve a destructive interference with SM,
since $C_{9\ell}^\text{SM}(m_b) = 4.2$~\cite{WC-SM}.
As can be seen from the figure, the fits do not determine the sign of
$C_{9e}^\text{NP}$, however they indicate $|C_{9e}^\text{NP}| < |C_{9\mu}^\text{NP}|$.

In an earlier paper~\cite{prev-paper}, we had identified a class of
minimal models that explain the flavor anomalies through the NP contributions to 
$C_{9\mu}^\text{NP}$ and $C_{9e}^\text{NP}$, in a bottom-up approach.
These models augmented SM by a $U(1)_X$ symmetry, which was instrumental in
generating the LFUV needed, and was broken spontaneously at the low scale
by an SM-singlet scalar $S$.
Three right-handed neutrinos helped generate neutrino masses through
the Type-I seesaw mechanism, with the same scalar $S$ instrumental in
obtaining the appropriate texture zeros that give rise to the observed
neutrino mixing pattern.
The number of particles beyond the SM was minimal -- apart from the $Z'$ gauge
boson associated with the $U(1)_X$, one only needed the scalar $S$ and an
additional Higgs doublet to generate quark mixing.
Appropriate $X$-charges were given to all particles such that the models are
anomaly-free, fermions charges are vector-like, and experimental constraints
from flavor physics --- in particular the negative sign of $C_{9\mu}^\text{NP}$
needed for explaining the $R_{K^{(*)}}$ anomaly --- were satisfied.
This class of models was consistent with all the experimental
measurements available at that time~\cite{rk-2014,utfit,atlas-dilepton-old}.
Indeed, even with the current data, the specific one-dimensional scenario in ref.~\cite{prev-paper}
predicting
$C_{9,\mu}^\text{NP} = 3 C_{9,e}^\text{NP}$ is quite close to the best fit,
while that with $C_{9,\mu}^\text{NP} = -3 C_{9,e}^\text{NP}$ also provides a very good
fit~\cite{matias-2021}, as can also be seen from figure~\ref{fig:1}.
Such scenarios correspond to the leptonic symmetry combinations $L_e \pm 3L_\mu$,
with unconstrained $L_\tau$.

In this article, we show that recent strong constraints on the mass and
coupling of the $Z'$ boson from collider
experiments~\cite{atlas-dilepton,cms-dilepton}  make the above models unviable, 
if they are minimal flavor violation (MFV)-like, i.e. if the mixing parameters
involved in the $B_d$ and $B_s$ sector are CKM-like.
However, if this requirement, imposed implicitly on the class of models
in~\cite{prev-paper}, is relaxed by a single parameter, a broader
class of non-MFV models emerges, which retains all the desirable properties
of the above models. Among them the scenarios with non-zero NP contributions to
$C_{9e}$ survive the strong collider constraints, while the scenarios with only NP contributions to $C_{9\mu}$ stay disallowed.
This new class of non-MFV models thus offers the most preferred candidates
for the solutions of the neutral-current flavor anomalies through a
$U(1)_X$ symmetry. We term these as ``frugal'' models, since the number of
particles beyond SM needed to complete these models are minimal.
Note that the number of additional particles in this model stays 
the same as that in ref.~\cite{prev-paper}.
    
Several papers~\cite{Buras:2013qja,Altmannshofer:2014cfa,Crivellin:2015mga,Celis:2015ara,prev-paper,Bonilla:2017lsq,Tang:2017gkz,Bian:2017xzg,Fuyuto:2017sys,King:2018fcg, Duan:2018akc, Allanach:2018lvl,Allanach:2019iiy, Altmannshofer:2019xda,less-MFV-crivillin,Baek:2020ovw,ben-b3-lmu, Davighi:2021oel,Bause:2021prv}
have focused on $U(1)_X$ models as the solutions to the $b \to s \ell \ell$ anomalies,
either in isolation or by combining them with some other well-motivated SM problems, 
like neutrino masses, dark matter, fermion mass heirarchy, etc.
With the current stringent colliders constraints~\cite{Greljo:2017vvb,atlas-dilepton,cms-dilepton,Allanach:2019mfl}, 
the models have increasingly focused their attention on the scenarios where the collider 
constraints can be minimized.
Examples of these include the models with only third generation of 
quarks charged under the new gauge symmetry~\cite{Allanach:2018lvl,Allanach:2019iiy,ben-b3-lmu,Bonilla:2017lsq},
and models with vector-like additional quarks charged under the new symmetry~\cite{Altmannshofer:2014cfa,Bonilla:2017lsq}. 
Some of the recent works have combined $U(1)_X$ symmetry and leptoquarks for
simultaneously explaning $b \to s \ell \ell$ anomalies and the muon $g-2$
discrepancy~\cite{Greljo:2021npi}.
In this manuscript, we follow the principle of frugality in adding new particles
to the SM, and identify a class of symmetries which can simultaneosuly explain 
the $b \to s \ell \ell$ anomalies and neutrino mixing.

The paper is organized as follows.
In section~\ref{sec:bottom-up}, we recap the bottom-up construction of the class of
$U(1)_X$ models that address the $b \to s \ell \ell$ anomalies, quark mixing, and
neutrino mixing pattern. In particular, we describe the algorithm
for assigning appropriate $X$-charges to particles, while obeying the
theoretical and experimental constraints.
In section~\ref{sec:constraints}, we discuss the constraints on the mass
and coupling of $Z'$ boson in these models from neutral meson mixing and 
collider data. In section~\ref{sec:MFV}, we show that the after incorporating the 
experimental constraints, the ``MFV-like'' models do not survive.
section~\ref{sec:NMFV} shows that the introduction of a single non-MFV parameter
allows a larger class of models to account for
the flavor anomalies, while being consistent with all available constraints. 
Section~\ref{sec:conclusions} summarizes our results, and concludes.

\section{Constructing $U(1)_X$ models in a bottom-up approach}
\label{sec:bottom-up}

In this section, we recap our bottom-up approach~\cite{prev-paper} to
identify models with a vector-like $U(1)_X$ symmetry that can explain the
$b \to s \ell \ell$ anomalies through the NP WCs
$C_{9\mu}^{\rm NP}$ and $C_{9e}^{\rm NP}$.
We denote the generic form of this symmetry as
\begin{equation}
  X \equiv \alpha_1 B_1 + \alpha_2 B_2  + \alpha_3 B_3
  + \alpha_e L_e +  \alpha_\mu L_\mu +\alpha_\tau L_\tau \; ,
\label{eqn:u1x}
\end{equation}
where $B_i$ denotes the $i^{\rm th}$ generation baryon number
and $L_j$ denotes the lepton number for j-type lepton.
The corresponding $X$-charges of fermions are listed in Table~\ref{tab:1}.
Note that in addition to the SM fermions, we also have three right-handed
neutrinos.

\begin{table}[h!]
\begin{center}
\begin{tabular}{|c|c|c|c|c|c|c|}
\hline
 Fields & ${u,d}$ & ${c,s}$  & ${t,b}$ & $e,\nu_e$ &$\mu,\nu_{\mu}$ & $\tau,\nu_{\tau}$ \\ \hline
 $X$ & $\alpha_1/3$ & $\alpha_2/3$  & $\alpha_3/3$ & $\alpha_e$  &$\alpha_\mu$ & $\alpha_{\tau}$   \\ \hline
\end{tabular}
\end{center}
\caption{\label{tab:1} The (vector-like) $X$-charges of fermions.}
\end{table}

The $SU(2)_L$ gauge symmetry of the SM forces the $X$-charges of particles
belonging to the same doublet to be identical.
The fermion $X$-charges are vector-like, which helps in anomaly cancellation,
and also ensures that the contribution from the NP axial-vector currents
vanishes, i.e. $C_{10\ell}^{\rm NP} = C_{10\ell}' =0$.
The anomaly cancellation in this case is simple and further leads to only one condition
\begin{equation}
  \alpha_1 + \alpha_2 + \alpha_3 + \alpha_e + \alpha_\mu + \alpha_\tau =0 \;. 
  \label{eq:anomaly1}
\end{equation}

Before analyzing the detailed quantitative constraints on the $\alpha$
parameters, desirable conditions on these parameters may be obtained
using the following considerations:
\begin{itemize}
\item The NP should not significantly affect the observables in neutral
  meson mixing, which have been found to match the SM predictions to a
  great precision.
\item The mass matrices of up-type and down-type quarks should be able to
  give rise to the appropriate Cabibbo-Kobayashi-Maskawa (CKM) matrix.
\item There should not be any massless goldstone bosons produced due to symmetry breaking.
\item The Type-I seesaw mechanism should yield the observed
  pattern of neutrino masses and mixing.
\item As indicated by global fits to the flavor anomaly data, the
  magnitude of NP coupling
  of electron should be smaller than that of muon.
\item The NP contribution $C_{9\mu}^{\rm NP}$ must have a negative sign.
\end{itemize}
We shall apply these conditions successively in the following subsections.

\subsection{$X$-charges of quarks and the CKM matrix}
\label{sec:quark-mixing} 

The origin of the CKM matrix is in the diagonalization of up-type quark
mass matrix $M_u$ and the down-type quark matrix $M_d$ by the bi-unitary
transformations
\begin{equation}
  M_u^{\rm diag} = V_{uL}^\dagger M_u V_{uR} ~,~
  M_d^{\rm diag} = V_{dL}^\dagger M_d V_{dR} ~.
  \label{eqn:diagonalize}
\end{equation}
The CKM matrix is then given by $V_{\text{CKM}} = V_{uL}^\dagger V_{dL}$.

In the presence of a new $U(1)_X$ gauge symmetry,
the flavor-changing neutral currents (FCNC) induced by the new gauge boson $Z'$
would affect the neutral meson mixings by giving additional tree-level contributions
to the box diagram in the SM.
We focus on the constraints from  CP violation in $K-\bar{K}$ mixing ($\epsilon_K$),
and the mass splitting as well CP-violation in $B_d-\overline{B}_d$ as well as
$B_s-\overline{B}_s$ mixing. We ignore constraints from $\Delta m_K$, since its value is 
dominated by long distance effects~\cite{Grinstein:2015nya}. We also do not incorporate
constraints from $D-\overline{D}$ mixing for the same reason.
The NP contribution may be calculated by
writing down the Lagrangian for left-handed d-type quarks
$D_L \equiv (d_L, s_L, b_L)^T$ in their mass basis: 
\begin{equation}
  {\cal L}_{Z'} = g_{Z'} \overline{D}_L [V_{dL}^\dagger \cdot \mathbb{X}_q \cdot
    V_{\rm d_L}] \gamma^\mu D_L Z'_\mu \;,
    \label{eqn:zprimeLag}
  \end{equation}
where $\mathbb{X} \equiv {\rm diag}(X_u, X_c, X_t) = {\rm diag}(\alpha_1/3,
\alpha_2/3,\alpha_3/3)$. As shown in ref.~\cite{prev-paper},
the mixing in the right-handed d-quark sector may be chosen to be small, so 
that the contributions due to right-handed currents stay subdominant.
The relevant matrix elements that control the dominant NP contributions in the
$K$, $B_d$ and $B_s$ systems may be written as
\begin{eqnarray}
  K: \quad [V_{dL}^\dagger \cdot \mathbb{X}_q \cdot V_{dL}]_{12}
  & = & (X_u - X_c) [V_{dL}]_{ud}^* [V_{dL}]_{us}
  + (X_t - X_c) [V_{dL}]_{td}^* [V_{dL}]_{ts} \; , \\
  B_d: \quad [V_{dL}^\dagger \cdot \mathbb{X}_q \cdot V_{dL}]_{13}
  & = & (X_u - X_c) [V_{dL}]_{ud}^* [V_{dL}]_{ub}
  + (X_t - X_c) [V_{dL}]_{td}^* [V_{dL}]_{tb} \; ,\\
  B_s: \quad [V_{dL}^\dagger \cdot \mathbb{X}_q \cdot V_{dL}]_{23}
  & = & (X_u - X_c) [V_{dL}]_{us}^* [V_{dL}]_{ub}
  + (X_t - X_c) [V_{dL}]_{ts}^* [V_{dL}]_{tb} \; ,
  \label{eq:Kmix}
\end{eqnarray}
where the unitarity of $V_{dL}$ has been used.
The choice $X_u =X_c$, and the small values of $[V_{dL}]_{td}$ and $[V_{dL}]_{ts}$, 
allow us to minimize the strong constraints from the $K$ sector and somewhat weaker
constraints from the $B_d$ and $B_s$ sectors. 
The condition $X_u = X_c$ also implies an underlying $U(2)^3$ flavor symmetry
present in the Lagrangian, which is broken only by the Yukawa
interactions~\cite{isidori-straub}.
This has also been referred to 
as ``less-minimal flavor violation"~\cite{less-MFV,less-MFV-crivillin}.
The additional choice $V_{uL}= {\mathbb I}$ (or equivalently, $V_{dL} = V_{CKM}$)
  made in ref.~\cite{prev-paper} makes the scenario ``MFV-like",
wherein the combinations of CKM elements contributing
to the mixing in the $B_d$ and $B_s$ sectors are the same as those in the SM. It also 
ensures that the NP contribution from the second term to $K-\overline{K}$ mixing is suppressed by
$|[V_{\rm CKM}]_{td} [V_{\rm CKM}]_{ts}| \sim {\cal O}(\theta_C^{~5})$, where
$\theta_C \approx 0.2$ is the Cabibbo angle.

We continue to use the condition $X_u = X_c$ in this paper. 
Later, for non-minimal scenarios, we will relax the condition $V_{dL} = V_{\rm CKM}$,
however the smallness of $[V_{dL}]_{td}$ will still be valid, keeping in mind
the stringent constraints from kaon oscillation data. 
 
The condition $X_u = X_c$ (i.e. $\alpha_1 = \alpha_3$) also impacts the structure of
the Yukawa matrices.
Since in our framework, the SM Higgs doublet is uncharged under $U(1)_X$,
the only nonzero elements in the SM Yukawa matrix can be  the three diagonal
elements and the off-diagonal elements in the first two generations.
This would force the $2$-$3$ and $1$-$3$ mixings in the CKM matrix to be zero.
In order to prevent this, the SM needs to be augmented with an additional 
doublet $\Phi_{\text{NP}}$ whose $X$-charge equals $X_u - X_t$, or equivalently,
$(\alpha_1-\alpha_3)/3$. 
This would result in the NP contribution to the Yukawa matrices of up-type and
down-type quarks in the form 
\begin{equation}
{\mathcal Y}_u^{\text{NP}} =   \begin{pmatrix}
0 & 0 & 0\\
0 & 0 & 0\\
\times & \times & 0\\
\end{pmatrix} , \quad  
{\mathcal Y}_d^{\text{NP}} =   \begin{pmatrix}
0 & 0 & \times \\
0 & 0 & \times \\
0 & 0 & 0\\
\end{pmatrix} \;,
\label{eq:Y1-matrices}
\end{equation}
where $\times$ denotes nonzero elements.
These off-diagonal elements give rise to the required mixing
in the 2-3 and 1-3 sector, to reproduce the CKM matrix~\cite{prev-paper}. 
These Yukawa matrices can now give rise to the mass matrices
\begin{equation}
  M_u = \frac{v}{\sqrt{2}} \left({\mathcal Y}_u^{\rm NP}  \cos\beta +  
    {\mathcal Y_u^{\rm SM}}  \sin\beta \right) \;, ~
    M_d = \frac{v}{\sqrt{2}} \left({\mathcal Y}_d^{\rm NP}  \cos\beta +  
    {\mathcal Y_d^{\rm SM}}  \sin\beta \right) \;,
\end{equation}
where $\langle v_{SM} \rangle = v \sin\beta$ and
$\langle v_{NP} \rangle = v \cos \beta$ are the vacuum expectation values of
the SM Higgs $\Phi_{\rm SM}$ and the NP Higgs $\Phi_{\rm NP}$, respectively.
The matrices $M_u$ and $M_d$ are diagonalized by the unitary matrices $V_{dL}$,
$V_{dR}$, $V_{uL}$ and $V_{uR}$, as shown in eq.~(\ref{eqn:diagonalize}).
Note that the requirement of $X_u = X_c$ is instrumental in generating the CKM
matrix with only one additional Higgs doublet.

\subsection{$X$-charges of scalars} 
\label{sec:scalar}

Among the two Higgs doublets $\Phi_{\rm SM}$ and $\Phi_{\rm NP}$, the former
is a singlet under $U(1)_X$, to ensure nonzero diagonal elements in
the flavor basis. The latter has an $X$-charge equal to
$(\alpha_1-\alpha_3)/3$, as seen above.

The absence of a massless pseudoscalar, which would be created due to
the breaking of a global $U(1)_A$ symmetry in the Lagrangian, necessitates
the introduction of an extra scalar $S$, which has the same $X$-charge
as $\Phi_{\rm NP}$~\cite{prev-paper}. It allows a term 
$S \Phi_{\text{NP}}^\dagger \Phi_{\text{SM}}$ in the scalar sector,
which yields a mass for the pseudoscalar after the breaking of $U(1)_X$
where $S$ gets a vacuum expectation value. 
The $X$-charge of $S$ also needs to be $X_S = (\alpha_1 -\alpha_3)/3$.

\subsection{$X$-charges of leptons and Neutrino mixing}
\label{sec:neutrino}

The global fits to neutral-current flavor anomalies and other $b \to s$ data
in the $\left( C_{9\mu}^{\text{NP}}, C_{9e}^{\text{NP}} \right)$ plane strongly
indicate $|C_{9e}^{\text{NP}}| <  |C_{9\mu}^{\text{NP}}|$, which indicate
$|\alpha_e| < |\alpha_\mu|$ as seen in figure~\ref{fig:1}. We therefore take this to be
one of the conditions on our model.
The value of $\alpha_\tau$ remains unconstrained from the current measurements.

We determine the $X$-charges of the leptons which can explain the 
patterns of neutrino mixing well. 
In particular, we desire that the leptonic mixing arises completely in the neutrino
sector, where the neutrino mass is generated by the Type-I seesaw mechanism.
However, since the $X$-charges of the three lepton generations are, in general,
different, it would not be possible to generate off-diagonal
elements in the neutrino mass matrix, which are needed for the large
neutrino mixing observed. 
In our model, this can be achieved  without the need for any
additional particle, but by using the interactions of the neutrinos with
the scalar $S$ that is already present~\cite{prev-paper}.
The terms contributing to neutrino mass are: 
\begin{equation}
\mathcal{L}_{\nu,\text{mass}} = \overline{\nu_L}_i [m^\nu_D]_{ij} {\nu_R}_j 
+ \overline{\nu^c_R}_i [m^\nu_R]_{ij} {\nu_R}_j + 
\overline{\nu^c_R}_i [y^\nu_R]_{ij} {\nu_R}_j  S(S^\dagger) + \text{h.c.}\; ,
\end{equation}
where $[m^\nu_D]$ is the Dirac matrix, $[m^\nu_R]$ is the Majorana mass
matrix of right-handed neutrinos, and $i,j$ are flavor indices.
The effective Marorana mass
matrix after the symmetry breaking becomes 
\begin{equation}
[M^\nu_R]_{ij} = [m^\nu_R]_{ij} + \dfrac{1}{\sqrt{2}} [y^\nu_R]_{ij} v_S \; ,
\end{equation}
where $v_S$ is the vacuum expectation value of $S$.
Thus the mass matrix $M^\nu_R$ gets off-diagonal elements,
which further lead to the mixing of left-handed neutrinos, through 
Type-I seesaw formula
\begin{equation}
  [m_\nu] = - [m^\nu_D] \cdot [M^\nu_R]^{-1} \cdot [m^\nu_D]^T \; .
\end{equation}

In order for the above neutrino mass matrix $[m_\nu]$ to reproduce the
observed neutrino mixing pattern, only certain texture-zero patterns of
$[M^\nu_R]$ are allowed~\cite{Grimus:2004hf, neutrino-textures}.
A subset of these patterns may be created by appropriate choices of
the values of $\alpha_e, \alpha_\mu, \alpha_\tau$, and $X_S$
\cite{prev-paper}. A further subset satisfies the requirement
$|\alpha_e| < |\alpha_\mu|$. 
The leptonic symmetries $(\alpha_e L_e + \alpha_\mu L_\mu + \alpha_\tau L_\tau)$
that satisfy all these criteria are:
\begin{itemize}
  \item $a\left(L_\mu - L_\tau \right)$ or $a L_\mu$, with $X_S = \pm a$,
  \item $a\left( L_e-3L_\mu + L_\tau \right)$ or
    $a\left(L_e \pm 3L_\mu - L_\tau\right)$, with $X_S = \pm 2a$.
\end{itemize}
Here $a$ is the overall multiplicative factor.

\subsection{Scenarios indicated by the bottom-up construction}

Inferring the $X$-charges of leptons from the allowed leptonic symmetries, 
and using  the conditions $\alpha_1 = \alpha_2$ and
$X_S = (\alpha_1 - \alpha_3)/3$, the $X$-charges of all the other particles are
fixed automatically
by demanding the theory to be anomaly free.
The $X$-charges of all leptons, in turn, are fixed up to an overall
multiplying factor $a$ as seen in the last section. We fix the normalization
by choosing $a$ so as to make $\alpha_\mu =1$. All the $U(1)_X$ scenarios thus 
determined are listed 
in Table~\ref{tab:categories}. We further categorize
them depending on their values of $X_S$ and $\alpha_e$. This ensures that
flavor constraints for scenarios belonging to the same category are identical.
Note that the categories A, B and C listed in table~\ref{tab:categories},
with negative $X_S$ values, are the same as given in~\cite{prev-paper}.
The category D from ref.~\cite{prev-paper} is not present in the current version
because of the imposition of $|\alpha_e| < |\alpha_\mu|$. In addition, have also 
included the categories AA, BB and CC with positive $X_S$ values. 
This inclusion completes the set of scenarios allowed by conditions in
sections~\ref{sec:quark-mixing},~\ref{sec:scalar}, \ref{sec:neutrino}.
Note that for the categories in each pair (A, AA), (B, BB), and (C, CC),
the leptonic symmetries are identical, but the sign of $X_S$ is different.

\begin{table}[t]
\def\arraystretch{1.2}
\begin{center}
\begin{tabular}{ |c|c|c|c|c|c|c|c|c|c|} 
 \hline
 Category & Scenario & $X_S$ & Leptonic symmetry & $\alpha_1$ & $\alpha_2$  & $\alpha_3$ & $\alpha_e$ & $\alpha_\mu$ & $\alpha_\tau$ \\ \hline 
 A & A1 & -1 &$L_\mu- L_\tau$ & $-1$ & $-1$ & $2$ & $ 0$ & $1$ & $-1$   \\
  & A2 & -1& $L_\mu$ & $-\frac{4}{3}$ & $-\frac{4}{3}$ & $\frac{5}{3}$ & $0$ & $1$ & $0$  \\ \hline 
 B& B1 &  $-\frac{2}{3}$&$L_e-3L_\mu + L_\tau$ & $-\frac{7}{9}$ & $-\frac{7}{9}$ & $\frac{11}{9}$ & $ -\frac{1}{3}$ & $1$ & $-\frac{1}{3}$  \\ 
 &B2 & $-\frac{2}{3}$&$L_e-3L_\mu - L_\tau$ & $-1$ & $-1$ & $1$ & $ -\frac{1}{3}$ & $1$ & $\frac{1}{3}$   \\ \hline
C & C1 & $-\frac{2}{3}$&$L_e + 3 L_\mu - L_\tau$ & $-1$ & $-1$ & $1$ & $ \frac{1}{3}$ & $1$ & $-\frac{1}{3}$   \\ \hline \hline
AA & AA1& 1 & $ L_\mu- L_\tau$ & ${1}$ & ${1}$ & $-{2}$ & $ 0$ & $1$ & $-1$   \\ 
  & AA2 & 1& $ L_\mu$ & $\frac{2}{3}$ & $\frac{2}{3}$ & $-\frac{7}{3}$ & $ 0$ & $1$ & $0$  \\ \hline 
 BB & BB1 & $\frac{2}{3}$ &$L_e-3L_\mu + L_\tau$ & $\frac{5}{9}$ & $\frac{5}{9}$ & $-\frac{13}{9}$ & $ -\frac{1}{3}$ & $1$ & $-\frac{1}{3}$  \\ 
 & BB2 & $\frac{2}{3}$&$L_e-3L_\mu - L_\tau$ & $\frac{1}{3}$ & $\frac{1}{3}$ & $-\frac{5}{3}$ & $ -\frac{1}{3}$ & $1$ & $\frac{1}{3}$   \\ \hline
CC & CC1 & $\frac{2}{3}$& $L_e + 3 L_\mu - L_\tau$ & $\frac{1}{3}$ & $\frac{1}{3}$ & $-\frac{5}{3}$ & $ \frac{1}{3}$ & $1$ & $-\frac{1}{3}$   \\ \hline 
 \end{tabular}
\end{center}
\caption{\label{tab:categories} 
  The scenarios indicated by our bottom-up construction, categorized by the charge
  $X_S$ and $\alpha_e$. Categories A, B and C, with negative $X_S$ values are the same as given in ref.~\cite{prev-paper},
  while we include the categories AA, BB and CC here, which have positive $X_S$ values. } 
\end{table}

\subsection{The sign of $C_{9\mu}^{\rm NP}$ and the sign of $X_S$}
\label{sec:signofC9}

The NP in our class of models influences $R_{K^{(*)}}$ primarily
through the Wilson coefficients $C_{9\ell}$.
The contributions to $C_{9\ell}^\prime$ are small due to the
small mixing angles in
$V_{dR}$ (see ref.~\cite{prev-paper} and section~\ref{sec:NMFV}). 
The tree-level contributions to $C_{10\ell}^{(\prime)}$ operators are zero.
They may arise due the $Z-Z^\prime$ mixing, and vanish in the small $Z-Z^\prime$
mixing limit.
The effective Hamiltonian relevant for the process $B \to K^{(*)} \ell \ell$ is
\begin{eqnarray}
\mathcal{H}_{\rm eff} &=& - 
\left( \frac{4 G_{F}}{\sqrt{2}} \frac{e^2}{(4\pi)^2}[V_{\rm CKM}]_{tb} [V_{\rm CKM}]_{ts}^* \;
C_{9\ell}^{\text{SM}} \right) \left( \overline{s_L}  \gamma^\mu b_L \right)
  \left( \overline{\ell}  \gamma_\mu  \ell \right) \nonumber \\
&&  - \left(\frac{X_S \, \alpha_\ell \, g^2_{Z^\prime}}{ M^2_{Z^\prime}}
  [{V_{dL}}]_{tb} [V_{dL}]_{ts}^*\right) 
\left( \overline{s_L}  \gamma^\mu b_L \right)
  \left( \overline{\ell}  \gamma_\mu  \ell \right) \; .
  \label{eqn:heff}
    \end{eqnarray}
Since $C_{9\ell} = C_{9\ell}^{\rm SM} + C_{9\ell}^{\rm NP}$, the above equation
is equivalent to 
\begin{eqnarray}
 C_{9\ell}^{\rm NP}
 &=&  \frac{4\sqrt{2} \, \pi^2 \,  g^2_{Z^\prime}\,
   }{ \, G_F \, M^2_{Z^\prime}\, e^2\, 
   }\cdot X_S \,\alpha_{\ell}\cdot \frac{[V_{dL}]_{tb}[V_{dL}]_{ts}^*}{[V_{\rm CKM}]_{tb} [V_{\rm CKM}]_{ts}^*}\;.
\label{eqn:C9NP}
 \end{eqnarray}
Note that the WCs have scale dependence, however the
qualitative inferences in this section do not change while running
from the scale $M_{Z'}$ to $m_b$. 
From eq.~(\ref{eqn:C9NP}), the two relevant Wilson coefficients are related 
by $C_{9e}^{\text{NP}}(m_b) = \alpha_e C_{9\mu}^{\text{NP}}(m_b)$.

From the global fits, we have seen that the sign of $C_{9\mu}^{\text{NP}}$
needed to explain the observed $b \to s$ anomalies has to be negative,
for the NP to destructively interfere with the SM, where
$C_{9\mu}^\text{SM}$ is positive~\cite{WC-SM}.
This leads to
\begin{equation}
  X_S \cdot \frac{[V_{dL}]_{tb}[V_{dL}^{*}]_{ts}}{
    [V_{\rm CKM}]_{tb} [V_{\rm CKM}]_{ts}^*} < 0 \; ,
\label{eqn:rk}
\end{equation}
i.e., either the charge $X_S$ is negative, or the ratio
\begin{equation}
  \mathcal{R}_{\text{mix}} \equiv \dfrac{[V_{dL}]_{tb}[V_{dL}^{*}]_{ts}}{
    [V_{\rm CKM}]_{tb} [V_{\rm CKM}]_{ts}^*} \;
    \label{eqn:rmix}
\end{equation}
 is negative. In ref.~\cite{prev-paper},
the assumption of $V_{uL}=\mathbb{I}$ led to
$V_{dL} = V_{\rm CKM}$, so that $\mathcal{R}_{\rm mix}$ was always unity.
As a result, 
only those symmetry combinations where $X_S < 0$ had been selected.
These are the categories A, B, C shown in Table~\ref{tab:categories}.
In this paper, we follow a generalized approach, without the assumption
$V_{uL}= \mathbb{I}$. This allows three additional categories, viz. AA, BB and CC, 
as shown in Table.~\ref{tab:categories}, where the sign of $X_S$ is positive.

\section{Experimental constraints} 
\label{sec:constraints}

We work in the limit where all additional NP particles apart from $Z^\prime$ are
decoupled, and determine constraints in the plane of
$\left( M_{Z^\prime}, g_{Z^\prime} \right)$.
The global fits already provide constraints
on these parameters from radiative, semileptonic, and leptonic B decays.
These parameters can be further constrained by collider searches and the neutral meson
mixing data, which we describe in the following subsections. 
The constraints from neutrino trident production are sub-leading for the
relevant mass-coupling range of $Z^\prime$~\cite{ben-b3-lmu}, and 
electroweak precision constraints can be evaded when $Z-Z'$ mixing
is taken to be small~\cite{Erler:2009jh} and the other NP particles are decoupled; these constraints are not included in
our analysis.

\subsection{Collider constraints}
\label{sec:collider}

\begin{figure} [t]
\begin{center}
\includegraphics[width=0.45 \textwidth]{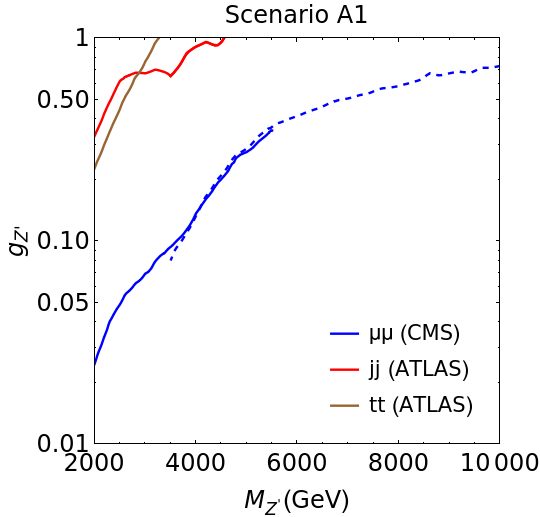}
\caption{Comparison of sensitivities of $t\bar{t}$~\cite{ATLAS:2020aln},
  dijet~\cite{ATLAS:2019fgd} and dimuon~\cite{cms-dilepton} resonance searches for
  the scenario A1 (which is likely to lead to the strongest $t\bar t$ bounds).  
  The limits from dimuon channel are much stronger than the other searches.
  The dimuon limits are extrapolated in the range
   5.5 GeV $ <M_{Z}^\prime \leq $ 10 TeV
  (dashed) using the shape of the dimuon invariant mass spectrum, and can be seen
  to match those from the available dimuon CMS search~\cite{cms-dilepton} in the
  interval 3.5 GeV $ \leq M_{Z}^\prime \leq$ 5.5 TeV.} 
\label{fig:compare-channels}
\end{center}
\end{figure}

The large amount of data being collected at the LHC strongly constrains 
any new physics that couples to light quarks.
Our scenarios in table~\ref{tab:categories} necessarily have non-zero couplings
of $Z^\prime$ to the first two generations of quarks.
Therefore, the $Z'$ particle will be produced
at the LHC for low $M_{Z^\prime}$ and high enough $g_{Z^\prime}$ values.
Furthermore, even a $Z^\prime$ that couples dominantly to the third generations
may be produced in a pp collision, albeit with a smaller cross section due to the
smaller parton fraction in the proton.
The non-observation of any such particle so far puts severe constraints on
model parameters.
The main observations that would constrain our class of models are:

\begin{itemize} 

\item Top-quark pair production limits from $pp \rightarrow Z^\prime \rightarrow t \bar t$~\cite{CMS:2020tvq, ATLAS:2020aln, ATLAS:2020lks} 

\item Dijet limits from $pp \rightarrow Z^\prime \rightarrow q \bar q$, including $b \bar b$~\cite{ATLAS:2019fgd}

\item Dilepton limit from $pp \rightarrow Z^\prime \rightarrow e^+ e^-,\, \mu^+ \mu^-$~\cite{atlas-dilepton,cms-dilepton} (including the non-resonant shape of the
  $m_{\mu\mu}$ distribution tail~\cite{altas-non-resonant, cms-non-resonant})
\end{itemize}
As we show below, the most stringent limits come from dimuon searches. A comparison of
the limits from all above observables can be seen in figure~\ref{fig:compare-channels}.

For $t \bar t$ searches, currently there is a measurement
of $t \bar{t}$ pair production cross section with 35.9 fb$^{-1}$ data
from CMS~\cite{CMS:2020tvq} and 139 fb$^{-1}$ data from ATLAS~\cite{ATLAS:2020aln}.
Limits can be calculated either by using the total cross section or
the invariant mass spectrum shape in addition.
The best measurement for the total cross section is currently
$830 \pm 39$~pb~\cite{ATLAS:2020aln},
and is completely consistent with the calculated SM cross section.
We therefore require that the contribution from NP to the total
  $t \bar t$ cross section keeps the prediction less than two sigma away from
the measured value.
Further incorporating the $t \bar{t}$ spectral shape, a lower bound
  $M_{Z'} \gtrsim 3.2 $ TeV is obtained for $g_{Z^\prime} = 1$, as can be seen in
figure~\ref{fig:compare-channels}.

For $M_{Z'} \lesssim 3$ TeV, the background for dijet searches is high as compared to
that for the $t \bar{t}$ searches, and hence the sensitivity of $t \bar{t}$ searches
is better. However, at higher $Z'$ masses, the dijet searches give slightly stronger
constraints. For example, $M_{Z'} \gtrsim 4.5$ TeV for $g_{Z^\prime} = 1$, as seen
in figure~\ref{fig:compare-channels}.

Our condition $|\alpha_e| < |\alpha_\mu|$ implies that the constraints from the
di-electron searches would always be weaker than those from the dimuon searches.
Hence, we focus on the dimuon channel. Experiments provide measurements of the
invariant mass spectrum in the dimuon final state. Due to the simplicity of the 
final state, this may be interpreted
in terms of a 95\% upper limit
on the production cross-section of $Z^\prime$ (with minimal fiducial cuts).
The parameter space ($M_{Z^\prime}, g_{Z^\prime}$) can then be constrained in any given
scenario by comparing the theoretical production cross-section with the experimental
95\% confidence limits. 
Such upper limits are available from the ATLAS experiment~\cite{atlas-dilepton}
for $M_{Z^\prime} <6 $ TeV,  and from the CMS experiment~\cite{cms-dilepton}
for $M_{Z^\prime} < 5.5$ TeV.
We use the constraints from CMS,
which are slightly stronger than those from ATLAS, to represent the dimuon limits.
As we see from figure~\ref{fig:compare-channels}, dimuon constraints are much stronger
than either $t \bar t$ or dijet constraints for the scenario A1 (or equivalently, AA1).
This observation remain true for all scenarios listed in table~\ref{tab:categories}.   

Using the number of observed events with high invariant mass,
we can extend the dimuon limits to higher values of $M_{Z^\prime}$ for which the
calculated limits have not been published by the experimental analyses
(see figure~\ref{fig:compare-channels}). 
The details of our calculations are explained in appendix~\ref{sec:appendix}.
The published bounds from CMS are available for $M_{Z'} \leq 5.5$ TeV.
Our calculated limits agree with these
limits in the region $ 3.5 \;{\rm TeV} \leq M_{Z^\prime} \leq 5.5$ TeV, 
thereby justifying the method used for extrapolation.
For the rest of this study, we use the published CMS dimuon limits upto 
$M_{Z^\prime } \leq 5.5 $ TeV, and our extrapolation for masses
$ 5.5 \;\text{TeV} <M_{Z^\prime} \leq 10$ TeV.

Note that a 10 TeV $Z^\prime$ can be excluded for high-enough coupling for all
of our scenarios.
On the other hand, requiring at least three events as a threshold for
detection puts the LHC reach for the discovery of $Z'$ to $M_{Z'} \approx 4-6$ TeV,
depending on the scenario in table~\ref{tab:categories}.

\subsection{Neutral meson mixing constraints}
\label{sec:mixings}

Apart from mediating tree-level
$b \to s \ell \ell$ transitions, the additional $Z^\prime$ particle would 
also be responsible for generating
tree-level mixing in $B_{d}-\overline{B}_{d}$, $B_s-\overline{B}_s$
and $K-\overline{K}$ sectors. 
These new physics contributions are heavily 
constrained from data~\cite{utfit}. Since the mixing
constraints are not taken into consideration in
global fits~\cite{matias-2019,mahmoudi-2019,matias-2021,mahmoudi-2021},
one has to incorporate them separately. Additionally, 
the new physics contributions generated by $Z^\prime$ only affect 
the operators with left handed quark currents, as the right handed mixing
is smaller in comparison to the left handed mixing (see ref.~\cite{prev-paper} and
section~\ref{sec:NMFV}). 
Hence $Z^\prime$ contributes to the same operators as in the SM. 
We get
\begin{equation}
 C^{\rm SM}_P(\mu)  \to  C^{\rm tot}_P(\mu) = C^{\rm SM}_P(\mu) + C^{\rm NP}_P(\mu) \; ,
 \end{equation} 
where $C^{\rm NP}_P$ at the $M_{Z^{\prime}}$ scale are given as
\begin{eqnarray}
C^{\text{NP}}_{K} (M_{Z^{\prime}})    &=&   
\frac{{2 \pi^2 \,X_S^2\,g^2_{Z^\prime} \left([V_{dL}]_{td}[V_{dL}]^{*}_{ts}\right)^2}}
 { M^2_{Z^\prime}G_F^2 M_W^2 } \;, \nonumber \\
C^{\text{NP}}_{B_d} (M_{Z^\prime})    &=& 
\frac{{2 \pi^2 \,X_S^2\,g^2_{Z^\prime} \left([V_{dL}]_{tb}[V_{dL}]^{*}_{td}\right)^2}}{ M^2_{Z^\prime}G_F^2 M_W^2 \left([V_{\rm CKM}]_{tb}[V_{\rm CKM}]^{*}_{td}\right)^2} \;, \nonumber \\
C^{\text{NP}}_{B_s} (M_{Z^\prime})    &=& 
\frac{{2 \pi^2 \,X_S^2\,g^2_{Z^\prime} \left([V_{dL}]_{tb}[V_{dL}]^{*}_{ts}\right)^2}}{ M^2_{Z^\prime}G_F^2 M_W^2
\left([V_{\rm CKM}]_{td}[V_{\rm CKM}]^{*}_{ts}\right)^2 } \;.
\end{eqnarray}
Here $P$ generically refers to one of the $B_d$, $B_s$ or $K$ meson.  
Note that while the CKM factors explicitly appear for $B-\overline{B}$ mixings,
they are conventionally absorbed in $C^{\text{SM}}_{K}(\mu)$.
After incorporating the running of the effective operators at one-loop order in QCD
at $M_W$ scale~\cite{buras-review}, the WCs are obtained as
\begin{equation}
C^{\text{NP}}_{P} (M_W)  = \left[\frac{\alpha_s(m_t)}{\alpha_s(M_W)}\right]^{\frac{6}{23}}\,
\left[\frac{\alpha_s(M_{Z^\prime})}{\alpha_s(m_t)}\right]^{\frac{2}{7}} 
C^{\text{NP}}_{P} (M_{Z^\prime}) \;,
\label{running}
\end{equation}
where $P$ stands for $K, B_d$, or $B_s$.
Note that the running of SM and NP is identical after the $M_W$ scale, hence we have
taken the running here only upto $W$-mass scale. These additional contributions
to $P-\overline{P}$ mixing get constrained from the measurements. The constraints
on $\Delta m$ and CP-violating phases are  parameterized~\cite{utfit} in terms of
\begin{equation}
C_{\epsilon_K} \equiv  \frac{\text{Im}\left[\left\langle K_0| \mathcal{H}^{\text{tot}}_{\text{eff}}|  
\bar{K_0}\right\rangle\right]} {\text{Im}\left[\left\langle K_0| \mathcal{H}^{\text{SM}}_{\text{eff}}|  
\bar{K_0}\right\rangle\right]} \;, \quad
C_{B_q} e^{2i\phi_{B_{q}}} \equiv  \frac{\left\langle B_q| \mathcal{H}^{\text{tot}}_{\text{eff}}|  
\bar{B_q}\right\rangle}{\left\langle B_q| \mathcal{H}^{\text{SM}}_{\text{eff}}| \bar{B_q}\right\rangle}\;,
\label{eqn:mixing-constraint}
\end{equation}
which can be studied in the plane of $(M_{Z^\prime}, g_{Z^\prime})$ for a given 
symmetry and a given $V_{dL}$.
Note that as mentioned in section~\ref{sec:bottom-up}, we do not consider the
constraint from  $\Delta m_K$ as it is dominated by long distance
corrections~\cite{utfit}.
For constraining our model parameter, we shall require that
the allowed  parameter space lies within 2$\sigma$ uncertainties 
for all these five observables, viz. $C_{\epsilon_K}$, $C_{B_d}$, $C_{B_s}$, $\phi_{B_{d}}$,
and $\phi_{B_{s}}$.

\section{Testing the scenarios against experimental constraints}
\label{sec:results}

In any given scenario, the flavor constraints crucially depend on 
$V_{dL}$. 
Indeed, as can be seen in eq.~(\ref{eqn:C9NP}), the value 
of $C_{9\mu}^{\rm NP}$ is related to $X_S$ through $\mathcal{R}_{\rm mix}$,
which depends on $V_{dL}$. In ref.~\cite{prev-paper}, we had chosen 
the MFV-like scenario 
$V_{dL} = V_{\rm CKM}$ and $X_u =X_c$, which gave rise 
to $\mathcal{R}_{\rm mix} =1$
for $B_d-\overline{B}_d$ and $B_s -\overline{B}_s$ mixing. 
In this paper, we will also allow more general scenarios for $V_{dL}$.

\subsection{``MFV-like'' scenarios with $V_{dL} = V_{\rm CKM}$ }
\label{sec:MFV}

\begin{figure} [htb!]
 \begin{center}
 \includegraphics[width=0.325\textwidth ]{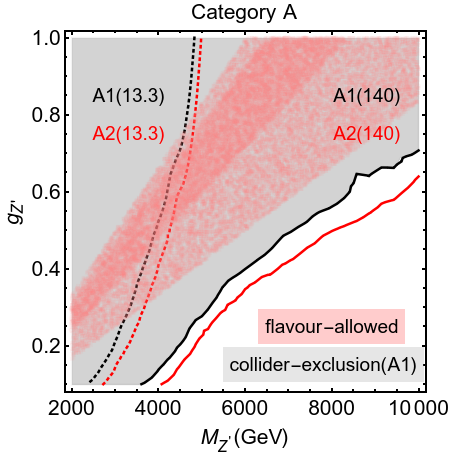}
 \includegraphics[width=0.325\textwidth ]{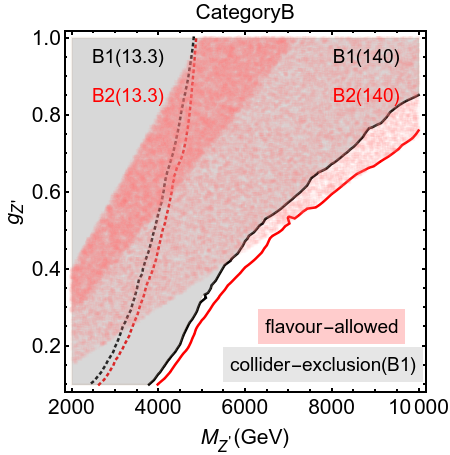}
 \includegraphics[width=0.325\textwidth ]{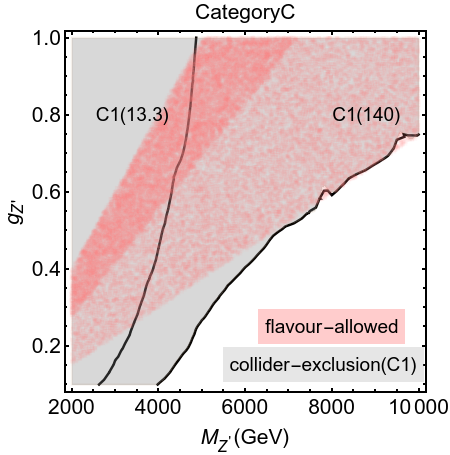}
 \caption{The constraints in the $\left(M_{Z^\prime}, g_{Z^\prime}\right)$ plane 
 for scenarios from categories A, B and C, with $V_{dL} = V_{\rm CKM}$.  
 While the light pink bands represent the combined $2\sigma$-allowed parameter space 
 from the meson mixing data~\cite{utfit16} and 
 $b \to s $ global-fit~\cite{mahmoudi-2016} in 2016,
 the darker bands include the 2018 constraints from the meson mixing data~\cite{utfit}
 and the 2021 updates to the $b \to s$ global fits~\cite{mahmoudi-2021}.
 The regions above the dotted (solid) lines are excluded at $95\%$ C.L.,
 with 13.3 (140) fb$^{-1}$ 
 total integrated luminosity, using dimuon
 searches~\cite{atlas-dilepton-old,cms-dilepton}. The gray 
 bands indicate the current exclusion for scenarios A1, B1 and C1.  }
 \label{fig:2}
 \end{center}
\end{figure}  

When $V_{dL}=V_{\rm CKM}$, the CKM factors in eq.~(\ref{eqn:C9NP}) cancel, and
the $C_{9\mu}^\text{NP}$ Wilson coefficient simplifies to
\begin{eqnarray}
 C_{9\mu}^{\text{NP}}(M_{Z^\prime}) 
 &=&  \frac{4\sqrt{2} \pi^2 X_S  \, g^2_{Z^\prime}}{G_F M^2_{Z^\prime} \, e^2}\;.
\label{eqn:mfvC9}
\end{eqnarray}
The desired negative value of $C_{9\mu}^{\rm NP}$ is obtained if $X_S <0$.
This points towards the scenarios belonging to the
categories A, B and C listed in table~\ref{tab:categories}.

We now subject these scenarios to the experimental constraints discussed
in section~\ref{sec:constraints}. 
The results are presented in 
fig~\ref{fig:2}. Note that for scenarios belonging to the same category, 
the global-fit constraints are identical, and so are the neutral 
meson mixings constraints. However collider constraints are different
for sub-scenarios (like A1 and A2)
which have different $X$-charge assignments for quarks. 
We can clearly see that, on one hand, 
the allowed $2\sigma$ bands from global-fit have started to become narrower, 
while on the other hand, the constraints from LHC are becoming considerably 
more stringent. The current data with 140 fb$^{-1}$ total integrated
luminosity~\cite{cms-dilepton} has essentially ruled out 
all the parameter space for these MFV-like models.

The freedom of choice of $V_{dL}$ allows us to find scenarios
that survive the stringent collider and meson-mixing constraints
above. This will be shown in the next subsection.

\subsection{Non-minimal flavor violating (non-MFV) scenarios }
\label{sec:NMFV}

Transition from MFV-like mixing, i.e. $V_{dL} = V_{\rm CKM}$, to
non-MFV mixing with $V_{dL} \neq V_{\rm CKM}$
would be severely constrained by measurements in the $K-\overline{K}$ sector, where 
the value of $\epsilon_K$ as given in eq.~(\ref{eqn:mixing-constraint}) is very well
measured. However, these constraints can be evaded if $V_{dL}$ is  chosen to be real.
In the rest of the paper, we shall continue with real $V_{dL}$.

As seen in section~\ref{sec:signofC9}, the resolution of $R_{K^{(*)}}$ anomalies
needs $X_S \mathcal{R}_{\text{mix}} <0 $.
Writing the real $V_{dL}$ in terms of three mixing angles $\theta_{12}$,
$\theta_{23}$, $\theta_{13}$ (similar to the CKM parameterization), 
$\mathcal{R}_{\text{mix}}$ in eq.~(\ref{eqn:rmix}) 
can be written as 
\begin{eqnarray}
\mathcal{R}_{\text{mix}} &=& \frac{[\cos\theta_{12} \cos\theta_{13} \sin{2 \theta_{23}}]_{dL}}
{[\cos\theta_{12} \cos\theta_{13} \sin{2 \theta_{23}}(1 + e^{-i \delta} \tan{\theta_{12}} \sin{\theta_{13}}\cot{\theta_{23}})]_{\rm CKM}}  \nonumber \\
&\approx& \quad \quad \quad \quad \quad \quad \frac{[\cos\theta_{12} \cos\theta_{13} \sin{2 \theta_{23}}]_{dL}}
{[\cos\theta_{12} \cos\theta_{13} \sin{2 \theta_{23}}]_{\rm CKM} }\;.
\label{eqn:refine-rmix}
\end{eqnarray}

Note that, since $\mathcal{R}_{\rm mix}$ can have either sign, the sign of $X_S$
can now be positive as well as negative. This allows the categories AA, BB and CC
from table~\ref{tab:categories} to be viable candidates, in addition to the
categories A, B and C considered earlier. Moreover, if 
the magnitude of $\mathcal{R}_{\rm mix}$ is large, 
the required values of $C_{9\mu}^{\rm NP}$ may become possible even with lower
values of $g_{Z^\prime}/M_{Z^\prime}$, as can be seen from eq.~(\ref{eqn:C9NP}).
However, the parameter $\mathcal{R}_{\rm mix}$ cannot be too large, otherwise the
simultaneous explanation of $b \to s \ell \ell$ anomalies along with neutral
meson mixing constraints from $B_{d/s}-\overline{B}_{d/s}$ mixing would be difficult.
Thus, a modest enhancement of $\mathcal{R}_{\rm mix}$ is required to make these scenarios
compatible with the global fits, neutral meson mixing data, and collider constraints.

Since $[\cos\theta_{12}  \cos\theta_{13}]_{\rm CKM} \approx 1$, one would need 
$[\sin 2\theta_{23}]_{dL} \gtrsim [\sin 2\theta_{23}]_{\rm CKM}$ for the
enhancement in $\mathcal{R}_{\rm mix}$.
In a simplified scenario, we can take
$\theta_{12,dL} \approx 0$ and $\theta_{13,dL} \approx 0$, which leads 
to 
\begin{equation}
\mathcal{R}_{\text{mix}} \approx  \frac{ [\sin{2 \theta_{23}}]_{dL}}
{ [\sin 2 \theta_{23}]_{\rm CKM} }\;.
\end{equation}
The choice of small $\theta_{12,dL}$ and $\theta_{13,dL}$ would also 
limit the severity of collider constraints. Note that our choice of $V_{dL}$
is the same as that in ref.~\cite{ben-b3-lmu}. 
This choice of $V_{dL}$ makes the constraints from $B_s-\overline{B}_s$ mixing
to be very crucial. 
From eq.~(\ref{eqn:diagonalize}), one can then obtain the corresponding matrix in $V_{dR}$  
as
\begin{equation}
\theta_{12,dR} \approx 0 \;, \quad \theta_{13,dR} \approx 0 \;, \quad 
\theta_{23,dR} = \tan^{-1}\left({\frac{m_s}{m_b} [\tan \theta_{23}]_{dL}}\right) \;.
\end{equation} 
It can be seen from this equation, that the mixing induced due to $V_{dR}$ remains small 
unless we are close the limit where $\theta_{23,dL} \to n \pi/2$.
We will stay away from these
limits in this paper. Our approximation of ignoring the right handed currents, used in eqns.~(\ref{eqn:zprimeLag},~\ref{eqn:heff}) and 
section~\ref{sec:mixings}, is thus justified. 

\begin{figure} [h!]
 \begin{center}
 \includegraphics[width=0.4\textwidth ]{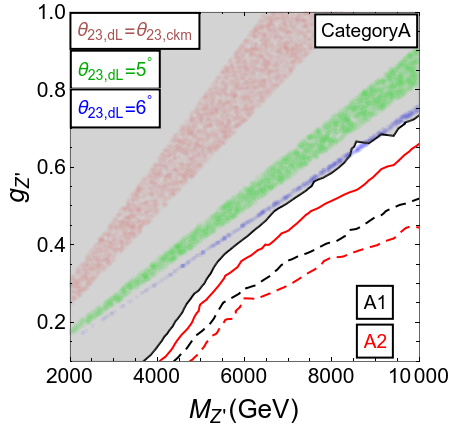}
 \includegraphics[width=0.4\textwidth ]{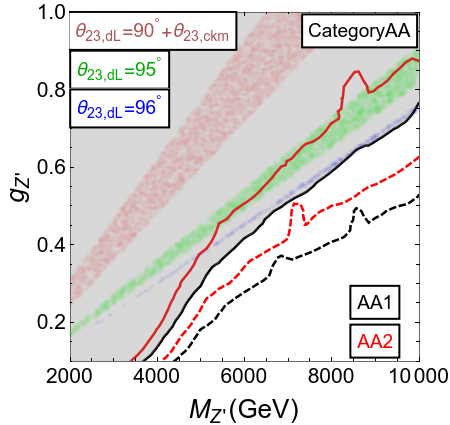}
 \includegraphics[width=0.4\textwidth ]{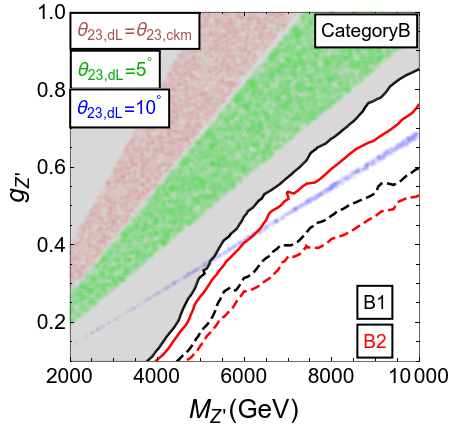}
 \includegraphics[width=0.4\textwidth ]{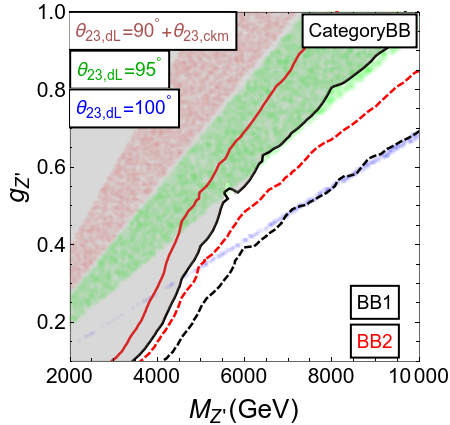}
  \includegraphics[width=0.4\textwidth ]{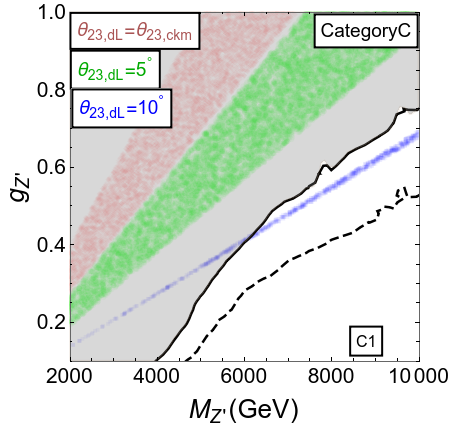}
 \includegraphics[width=0.4\textwidth ]{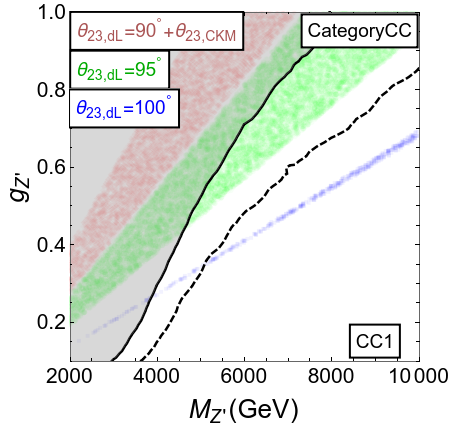}
 \caption{
    The constraints in the $\left( M_{Z^\prime}, g_{Z^\prime}\right)$ plane for the 
 non-MFV scenarios, for fixed values of $\theta_{23,dL}$. 
 The pink, green, and blue coloured bands indicate the combined $2\sigma$-allowed regions
 from the $b \to s$ global fit~\cite{mahmoudi-2021} and neutral meson mixing
 data~\cite{utfit}.
 The regions above the solid lines are excluded to $95\%$ C.L., with 140 fb$^{-1}$ 
 total integrated luminosity, using dimuon searches~\cite{cms-dilepton}. The gray 
 bands indicate the current $95\%$ C.L. exclusion regions for scenarios A1, B1, C1, AA1, BB1 
 and CC1, in the respective plots.
 The dashed lines represent the reach 
 of collider constraints with 3000 fb$^{-1}$ total integrated luminosity.  }
 \label{fig:4}
 \end{center}
\end{figure}  

The introduction of non-minimal flavor violation in its frugal form has allowed 
us an extra parameter $\theta_{23,dL}$.
The sign of $C_{9\mu}^{\text{NP}}$ dictates that the symmetries
in categories A, B and C will work if $\theta_{23,dL}$ is in the first quadrant,
and categories AA, BB and CC will work if $\theta_{23,dL}$ lies in the second quadrant. 

In figure~\ref{fig:4}, we present the main results of this section in the plane
of $\left( M_{Z^\prime}, g_{Z^\prime}\right)$, for a few selected values of $\theta_{23,dL}$. 
From the figure, the following observations may be made: 
\begin{itemize}

\item  For a given category, the combined constraints from the $b \to s$ global fit and
  neutral meson mixing with a given $\theta_{23,dL}$ value are identical to those
  with $90^\circ-\theta_{23,dL}$.
  
\item As the flavor constraints depend on $X_S \mathcal{R}_{\rm mix}$, they can be 
  identical for the scenarios that have the same value of $|X_S|$ but opposite sign,
  with $\theta_{23,dL}$ values differing by $90^{\circ}$.
  For example, compare A($\theta_{23,dL} = 5^{\circ}$) with AA($\theta_{23,dL} = 95^{\circ}$).
 The collider constraints for these pairs are, however, different.

\item The constraints for B2 and C1 are almost identical, and so are the constraints 
for BB2 and CC1. This is because the scenarios in these pairs carry identical $X$ charges for quarks and muons.
They differ only in the sign of $X_e$, however the global fit~\cite{mahmoudi-2021}
is nearly symmetric in $C_{9e}^{\rm NP}$, as can be seen in figure~\ref{fig:1}.
 
\item In the categories A, B, C, smaller $\theta_{23,dL}$ values
  $\approx 5^\circ - 10^\circ$ satisfy the flavor constraints,
  neutrino mixing, and collider constraints simultaneously.
   However, this is not possible for larger $\theta_{23,dL}$ values, as may be seen
  from the thinning of the colored bands with an increase in $\theta_{23,dL}$.
  This happens because $C_{9\ell}^{\rm NP}$ is proportional to $\mathcal{R_{\rm mix}}$,
  while the $B_s-\overline{B}_s$ mixing is sensitive to $\mathcal{R}^2_{\rm mix}$, and
  does not allow it to take a larger value.   
  A similar comment applies to the categories AA, BB, CC, where the allowed
  $\theta_{23,dL}$ values are $\approx 95^\circ - 100^\circ$.

\item The symmetries belonging to categories A and AA, where new physics
  contributes only in the muon (and/or) tau sector, stay ruled out from the
  current constraints on the dimuon resonance search at LHC~\cite{cms-dilepton}.
  At higher luminosities of 3000 fb$^{-1}$ at the LHC, the parameter space relevant for
  scenarios B1, B2, C1 and BB1 may also be completely probed by the collider searches,
  for $M_{Z^\prime} \leq 10$ TeV.

\item The scenarios BB2 and CC1 will be the most difficult to rule out even 
  with the high luminosity run of the LHC.
  This is expected since they have the smallest $X$-charges for quarks among all
  the categories (see table~\ref{tab:categories}).
  These scenarios correspond to the leptonic symmetry combinations
  $L_e \pm 3 L_\mu+L_\tau$, with positive $X_S$ values.
  
  \end{itemize}

\begin{figure} [t]
  \begin{center}
    \includegraphics[width=0.48\textwidth ]{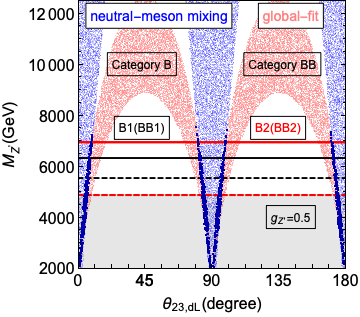}\;\;
    \caption{The constraints in the plane of $\left( \theta_{23,dL}, M_{Z^\prime}\right)$
      for a fixed value of $g_{Z^\prime}$, for categories B and BB. 
      The pink and blue bands show the $2\sigma$-allowed regions from the
      $b \to s$ global fit~\cite{mahmoudi-2021} and neutral meson mixing
      data~\cite{utfit}, respectively. The thin purple region satisfies
      the two constraints simultaneously.
      The regions below the solid (dashed)
      horizontal lines for
      scenarios in category B (BB) are excluded to 95$\%$ C.L., with 140 fb$^{-1}$ total integrated
      luminosity, using dimuon searches~\cite{cms-dilepton}.
      The gray band shows the current exclusion for the BB2 scenario.  }
    \label{fig:5}
  \end{center}
\end{figure}  

In figure~\ref{fig:5}, we show the incompatibility of the $B_s-\overline{B}_s$
constraints with the $b \to s$ global fit at large $[\sin 2\theta_{23}]_{dL}$ values,
for categories B and BB as representative examples. 
All our findings from figure~\ref{fig:4} may be reconfirmed here.
The neutral-meson mixing constraints for the pairs of categories 
(B, BB) are identical and the global fit constraints are mirror images of
one another around $90^\circ$.
Only the tiny narrow regions, shaded in purple, survive both these simultaneously.
It can also be noted that the collider constraint is the weakest for the scenario BB2.
Hence, this scenario is expected to be the most difficult to rule out even with the
higher luminosity runs of LHC.  

Indeed, for the scenarios with first two generations of quarks charged under $U(1)_X$,
it is difficult to simultaneously explain the $b \to s \ell \ell$ anomalies
along with neutrino mixing and neutral meson mixing, while staying  
compatible with the collider constraints.
In this section, we identified a suitable simple choice of $V_{dL}$ that can circumvent 
the otherwise stringent collider constraints for some of the scenarios, without the addition 
of any new particle in our construction.
Even with this non-minimal flavor violation, the scenarios with leptonic symmetry
combinations $L_\mu-L_\tau$ and $L_\mu$ stay ruled out.
The leptonic symmetry combinations $L_e \pm 3 L_\mu - L_\tau$ and
$L_e - 3 L_\mu + L_\tau$ emerge as the viable ones with the current data,
though they will be further probed with the high-luminosity data at the LHC,
with 3000 fb$^{-1}$ of integrated luminosity.
Thus in our frugal setup, the data seems to hint towards the possibility of
new physics in the electron as well as tau sector, in addition to the muon sector.

\section{Summary and concluding remarks}
\label{sec:conclusions}
  
In our present work, we identify a class of $U(1)_X$ models which 
can simultaneously explain the $b \to s \ell \ell$ anomalies and neutrino
mixing patterns.
We identify the $X$-charges 
using hints from the previous measurements and global fits in a bottom-up approach.
We follow  
the principle of frugality, i.e., try to  minimize 
the number of additional fields beyond SM. The only fields added are three
right-handed neutrinos, an additional SM doublet Higgs, and a SM-singlet scalar. 
The methodology followed here is similar to the one considered in ref.~\cite{prev-paper}.

We focus on the construction of scenarios where the NP contributes primarily to
$\mathcal{O}_{9\mu}$ as well as $\mathcal{O}_{9e}$. 
The global fits~\cite{matias-2021,mahmoudi-2021} imply the sign of $C_{9\mu}^{\rm NP}$
has to be necessarily negative, and the magnitude of new physics contributions in
electron has to be smaller than muon. The sign of $C_{9e}^{\rm NP}$ is not constrained 
by the global fits. 
The choice of vector-like $X$-charges ensures vanishing $C_{10}^{(\prime)}$, and helps
make the theory anomaly-free.
Note that contributions due to $\mathcal{O}_{9\ell}^{'}$ also 
remain negligible in our analysis.

The stringent constraint from $K-\overline{K}$ implies equal charges for the first
two quark generations. 
The requirement of generating 
$\overline{b}\gamma_\mu P_L s {Z^{\prime}}^\mu$ interaction through tree-level exchange 
of $Z^\prime$ dictates that the $X$-charge of the third generation quarks must be necessarily different
from the first two. 
The additional Higgs doublet with an appropriate $X$-charge then generates the 
desired quark mixing.

The singlet scalar $S$ breaks the $U(1)_X$ symmetry spontaneously and helps
generate the neutrino masses and their mixing pattern.
The choice of equal $X$-charges of $S$ and $\Phi_{\rm NP}$ prevents
the emergence of a massless Goldstone boson in the spectrum.
This also relates the $X$-charges of quarks with the leptons, 
which can be uniquely determined
using the requirement of anomaly cancellation. 

The observed neutrino mixing patterns
restrict the possible leptonic symmetries in our frugal set-up, 
where the scalar singlet $S$ is sufficient to generate the neutrino masses and
mixing patterns.  
This also leads to an important consequence that all the identified scenarios
necessarily have non-zero $X$-charges for all generations 
of quarks. This may be contrasted with the scenarios where only third 
generation of quarks are charged, e.g. $B_3 - L_\mu$ symmetry.
Such scenarios would require more particles than those that are already present
in our frugal set-up, for simultaneous explanations of   
neutrino mixing patterns and $b \to s \ell \ell$ flavor anomalies.

To generate the correct (negative) sign of $C_{9\mu}^{\rm NP}$, we find that the 
combination $X_S \mathcal{R}_{\rm mix}$ should be negative. In 
ref.~\cite{prev-paper}, where the MFV-like mixing $V_{dL} = V_{\rm CKM}$ was 
chosen, we had $\mathcal{R}_{\rm mix} =1$, which implied that 
only the scenarios with $X_S <0 $ can explain the flavor anomalies well.
However, allowing the departure
of $V_{dL}$ from $V_{\rm CKM}$ enables us to select a broader set 
of scenarios with both positive and negative signs of $X_S$.
In our analysis, we work in the limit where all additional NP particles apart from $Z^\prime$ are 
decoupled, so that the relevant parameter space is that of 
the mass and coupling of $Z^\prime$, viz. ($M_{Z^\prime}, g_{Z^\prime}$) for 
different choices of $V_{\rm dL}$.

Experimental limits from the collider searches 
and neutral meson mixing give the dominant constraints on ($M_{Z^\prime},g_{Z^\prime}$).
The neutral meson mixing constraints are evaluated for $K-\overline{K}$ and
$B_{d/s}-\overline{B}_{d/s}$ oscillations.
We compare the exclusion limits from resonance searches in dijet, $t \bar t$ and
dilepton channels, and find that the CMS dimuon search gives the most stringent
constraints for all scenarios.  We find that, after taking into account the recent
full run-2 data from the LHC, 
no MFV-like scenario compatible with the flavor anomalies remains allowed. 
The stringent collider constraints arise because of the non-zero $X$-charge
assignment of the first two quark generations, necessitated in our frugal set-up. 

By relaxing the assumption of the CKM-like 
mixing for $V_{dL}$, the collider constraints can be made compatible 
with the flavor anomalies for scenarios with leptonic symmetries of the form
$L_e \pm 3 L_\mu - L_\tau$ and $L_e - 3 L_\mu + L_\tau$. 
We demonstrate this with a simple non-MFV scenario where $V_{dL}$
only involves mixing between the second and the third generations, 
parameterized by $\theta_{23,dL}$. 
In order to generate the desired sign of $C_{9\mu}^\text{NP}$,
the new mixing angle $\theta_{23,dL}$ necessarily lies in the first (second) quadrant
for scenarios with negative (positive) $X_S$.
Note that scenarios with NP contributions present only in muon (and/or tau),
stay ruled out even when the mixing is allowed to be non-MFV.

We extrapolate the resonant dimuon search limits to $M_{Z^\prime}$ 
values upto 10 TeV,
to investigate future prospects for a $Z^\prime$ discovery.
 While the scenarios with leptonic symmetry
$L_e \pm 3 L_\mu -L_\tau$ and negative $X_S$, as well as 
 $L_e - 3 L_\mu + L_\tau$ with either sign of $X_S$, will be completely probed with
 3000 fb$^{-1}$ total integrated luminosity, the scenarios with $L_e \pm 3 L_\mu -L_\tau$
 and positive $X_S$ will be difficult to rule out even with the high luminosity run
 at the LHC.

To conclude, our class of frugal $U(1)_X$ models, that employ 
a minimal number of particles beyond the SM,
can account 
for $b \to s $ anomalies as well as the neutrino mixing pattern. 
The recent stringent collider constraints can be overcome
by a one-parameter 
choice of $V_{dL}$, without any additional particles,
for a set of scenarios where $Z'$ couples to all three lepton generations.

\section*{Acknowledgments}
The authors would like to thank Abhaya Kumar Swain and IACS for help with computational resources. 
A.D. acknowledges support of the Department of Atomic Energy (DAE), Government of India, under Project Identification No. RTI4002. ND acknowledges support from the Department of Science and Technology through grant number SB/S2/RJN-070.

\appendix
\section{Extrapolation of exclusion limits from dimuon searches}
\label{sec:appendix}

Here we describe the procedure for extrapolating the exclusion limits from collider searches, using the ATLAS dimuon search~\cite{atlas-dilepton} as an example.  In general, experiments provide the observed invariant mass spectrum in each final state.  This final shape depends on the production cross section, branching fraction for the relevant decay mode, as well as the detector acceptances and efficiencies. 
For a more complicated observable, it would be difficult for a phenomenological study to use this information without detailed description of the efficiencies. However, in the dimuon case, once the basic fiducial cuts (described in ref.~\cite{atlas-dilepton}) are taken into account, we find that we can reproduce the
published experimental limits accurately.

\begin{figure} [t]
\begin{center}
\includegraphics[width=0.6 \textwidth]{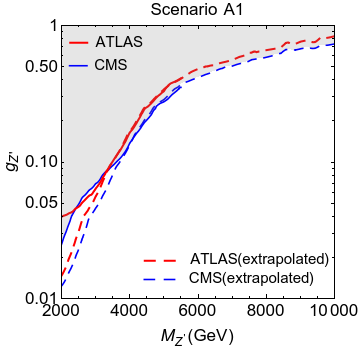}
\caption{Our calculated dimuon limits for scenario A1 in the $\left(M_{Z^\prime}, g_{Z^\prime}\right)$
plane, using the procedure described in the text. The comparison with the published 
results of ATLAS~\cite{atlas-dilepton} and CMS~\cite{cms-dilepton} collaboration shows that
our extrapolated results (dashed) agree with the published limits (solid) 
for high values of $M_{Z^\prime}$.
The matching for $M_{Z^\prime} \gtrsim 3.5$ TeV  for other scenarios in table~\ref{tab:categories} is at a similar level.
The gray shaded band in the figure highlights the full constraint used, i.e.~published limits from~\cite{cms-dilepton}
upto $M_{Z^\prime} < 5.5$ TeV and the extrapolated limits in the range 5.5 TeV $<M_{Z^\prime} <$ 10 TeV. } 
       \label{fig:compare-excl}
       \end{center}
       \end{figure}  

The total number of events expected from a signal hypothesis (choice of charges, $m_{Z^\prime}$, and $g_{Z^\prime}$) can be obtained by
\begin{equation}
N_\mathrm{sig} = \sigma_\mathrm{gen} \times \epsilon_\mathrm{fid} \times \mathcal{L_{\rm int}} \;,
\end{equation} 
where $\sigma_\mathrm{gen}$ is the production cross section into the dimuon final state, $\epsilon_\mathrm{fid}$ is the efficiency of the fiducial cuts, and $\mathcal{L}_{\rm int}$ is the integrated luminosity.  

A simple Poisson likelihood can be constructed using binned data.
In our case, since we are only interested in the extrapolation to high masses, we simplify the problem by looking at only the last bin, which collects all observations
with the dilepton invariant mass $M_{\ell \ell} > 2$~TeV for the ATLAS dilepton
search~\cite{atlas-dilepton}.
For a single bin, Poisson likelihood ($\mathcal{L}$) for observed number of events $n$
and expected number of events $\mu$ is given as
\begin{equation}
\mathcal{L}_\mu =  \frac{e^{-{\mu}} \mu^{n}}{{n}!} \;,
\end{equation}
leading to
\begin{eqnarray}
\chi^2 & = &  -2 \log \left( \frac{\mathcal{L}_{s+b}}{\mathcal{L}_b} \right) 
 =   -2   \left[ -s +   n \log \left( 1 + \frac{s}{b} \right) \right]\;.
\end{eqnarray}
Here $s$ and $b$ are the expected number of signal and background events, respectively. In order to get 95\% confidence limits,  the above equation is solved for $s$ for $\chi^2 = 3.841$, which corresponds to one-sided p-value of 0.05 for one degree of freedom.

The choice of only one bin for the high mass tail of the resonance mass distribution implies that 
our upper limits are conservative, since we do not use the information on the modification of the shape of the distribution due to non-resonant contribution.  Figure~\ref{fig:compare-excl} shows that this prescription matches the high-end ($M_{Z^\prime} > 3.5$ TeV) official limits very well, and therefore can be used reliably. This simple formulation thus allows us to extrapolate the limits for $M_{Z^\prime}$ up to 10 TeV, as well as to calculate expected sensitivities from future runs at the LHC (assuming that all events scale with integrated luminosity).  

A similar exercise may be carried out with the CMS data as well. However, we find that a flat overall efficiency factor of 0.4 is needed to reliably get the same upper limits as published by CMS~\cite{cms-dilepton} where the last bin collects all events with dimuon invariant mass greater than $ 1.8$ TeV. The comparison of our calculation with published ATLAS and CMS are shown in figure~\ref{fig:compare-excl}. Even though our calculated CMS limits deviate from the ones published with full shape analysis for low $M_{Z^\prime}$ as expected, they match very well for all values of $M_{Z^\prime} > 3.5$ TeV, and can be used for extrapolation to high masses with confidence.

 
\providecommand{\href}[2]{#2}\begingroup\raggedright\endgroup

\end{document}